\documentclass{hotnets2020}

\usepackage{blindtext}
\usepackage{color}
\usepackage{times}
\usepackage{xspace}
\usepackage{textcomp}
\usepackage{subcaption}
\usepackage{amssymb} 
\usepackage{pifont}
\usepackage{flushend}
\usepackage[hyphens]{url}

\usepackage[title]{appendix}

\usepackage{hyperref}

\newcommand{\jc}[1]{{\color{blue}{\footnotesize [JC: #1]}}}

\newcommand{\ignore}[1]{{\xspace}}

\newcommand{\video}{{visual rendering}\xspace}
\newcommand{\videor}{{visual renderer}\xspace}
\newcommand{\Video}{{Visual rendering}\xspace}
\newcommand{\videos}{{visual renderings}\xspace}
\newcommand{\Videos}{{Visual renderings}\xspace}
\newcommand{\vr}{{360\textdegree}\xspace}

\newcounter{packednmbr}

\newenvironment{packeditemize}{\begin{list}{$\bullet$}{\setlength{\itemsep}{0.5pt}\addtolength{\labelwidth}{-4pt}\setlength{\leftmargin}{2ex}\setlength{\listparindent}{\parindent}\setlength{\parsep}{1pt}\setlength{\topsep}{2pt}}}{\end{list}}

\newcommand{\tightcaption}[1]{\vspace{-0.3cm}\caption{{\normalfont{\textit{{#1}}}}}\vspace{-0.4cm}}
\newcommand{\tightsection}[1]{\vspace{-0.2cm}\section{#1}\vspace{-0.07cm}}

\newcommand{\tightsubsection}[1]{\vspace{-0.2cm}\subsection{#1}\vspace{-0.07cm}}

\newcommand{\eg}{{\it e.g.,}\xspace}
\newcommand{\ie}{{\it i.e.,}\xspace}

\newcommand{\mypara}[1]{\vspace{0.05cm}\noindent{\bf {#1}:}~}

\newcommand{\myparaq}[1]{\smallskip\noindent{\bf {#1}?}~}

\begin{document}
\title{A New Abstraction for Internet QoE Optimization}

\author{Junchen Jiang~~~~~~~~~~~~~~~~~~Siddhartha Sen\\
University of Chicago~~~~~~~~~Microsoft Research~~
}


\maketitle

\begin{abstract}
A perennial quest in networking research is how to achieve higher
quality of experience (QoE) for users without incurring more resources.
This work revisits an important yet often overlooked piece of the puzzle: what should
the {\em QoE abstraction} be? A QoE abstraction is a representation of
application quality that describes how decisions affect QoE.
The conventional wisdom has relied on developing handcrafted quality metrics (\eg video
rebuffering events, web page loading time) that are specialized to each application,
content, and setting. 
We argue that in many cases, it maybe fundamentally hard to capture a user's perception of
quality using a list of handcrafted metrics, and that expanding the metric list 
may lead to unnecessary complexity in the QoE model without a commensurate gain.
Instead, we advocate for a new approach based on a new QoE abstraction called {\em
\video}. 
Rather than a list of metrics, we model the process of quality perception as a user
watching a continuous ``video'' (\video) of all the pixels on their screen.
The key advantage of \video is that it captures the full experience of a user with the
same abstraction for all applications.
This new abstraction opens new opportunities (\eg the possibility of end-to-end deep learning models that infer QoE
directly from a \video) but it also gives rise to new research challenges 
(\eg how to emulate the effect on \video of an application decision).
This paper makes the case for \video as a unifying abstraction for Internet QoE and
outlines a new research agenda to unleash its opportunities.

\if
Traditionally, QoE optimization is abstracted to the optimization of a handful of objective quality metrics and much effort has been made to handcraft quality metrics.
We argue, however, that it is fundamentally hard to full capture the QoE perception process using just a series of quality metric numbers.
So adhering to quality metrics as QoE abstraction will either yield inaccurate QoE estimation (thus limiting QoE optimization), or even more handcraft quality metrics and more complex QoE models to capture various QoE-related factors, if these factors could ever be enumerated as more web and video applications run on diverse platforms and serve richer content.

In contrast, we advocate for a remarkably different abstraction of QoE optimization. 
We argue that QoE should be modeled by {\em \video}, which is a continuous video stream encompassing all pixels on the screen seen by a user.
In theory, \video captures the full experience of a user with the same abstraction for all applications. 
It is also amenable to computer vision-inspired techniques to infer QoE in an ``end-to-end'' fashion directly from what a user sees (much as convolutional neural networks revolutionized computer vision by replacing feature engineering with end-to-end training).
This paper makes the case for \video as a unifying abstraction of Internet QoE and outline a new research agenda to unleash new opportunities inspired by advances in computer vision (and visual cognitive research) which would be impossible in the conventional approaches.
\fi

\end{abstract}

\tightsection{Introduction}

An inflection point in Internet traffic is afoot, driven by a confluence of trends in
Internet applications (web services, video streaming, etc.): more devices with
larger screens, more high-fidelity content, more interactive applications, and more
impatient users~\cite{video-trends,Sandvine-trends}.
These trends are playing out against a backdrop of plateauing improvement in video
and web service quality despite considerable academic and industrial research
effort. The consequence is far reaching: application quality continues to fall
short of user expectations, and application demands increasingly overwhelm the
Internet's capacity---\eg content providers are forced to reduce streaming video quality
to cope with more users staying at home~\cite{quality-drop}.
These unprecedented challenges call for a new approach to achieving
higher quality of experience (QoE) for users given limited network resources.


The trade-off between QoE and resources has been widely studied in the networking and
multimedia communities.
A key concept underpinning most QoE optimization work is a {\em QoE model} that infers the
quality perceived by a user when interacting with an application, such as watching a
streaming video (\eg~\cite{balachandran2013developing,dobrian2011understanding}) or
loading a web page
(\eg~\cite{da2018narrowing,gao2017perceived,balachandran2014modeling,bocchi2016measuring}).
QoE models are integral to protocols/control algorithms that adapt their
decisions to maximize QoE given limited, dynamic availability of network resources
(bandwidth, latency, etc.).
An accurate QoE model enables these protocols to balance between conflicting
metrics (\eg when does a user prefer higher resolution over less rebuffering?) and
minimum resources to achieve high QoE (\eg for a given web page, how
fast is fast enough for users?).

\begin{figure}[t]
\centering
\includegraphics[width=1\columnwidth]{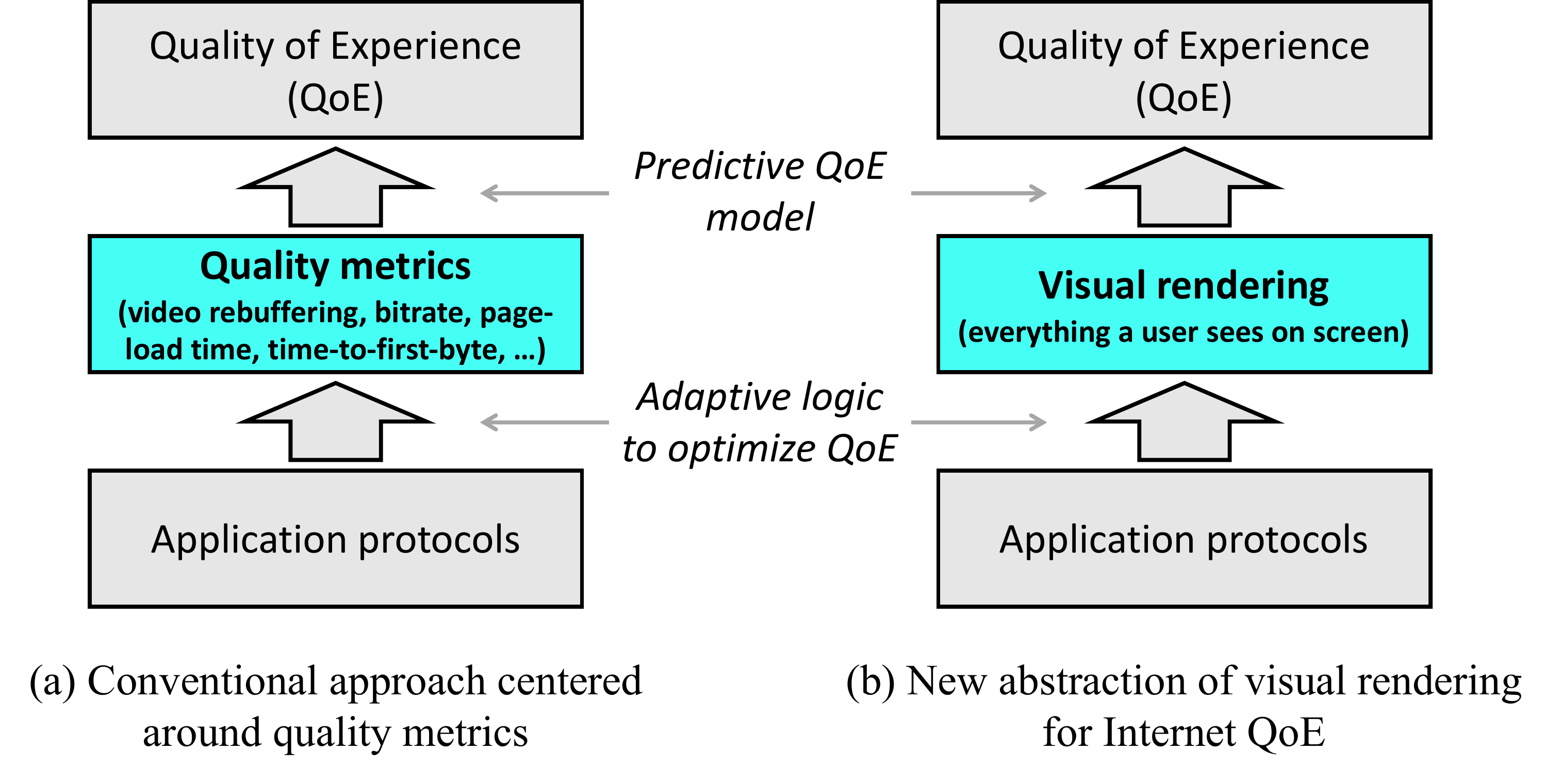}
\tightcaption{Traditional QoE optimization is based on quality metrics, which fail to
capture the full experience of a user. We propose a new abstraction called \video,
which captures user experience as a stream of pixels on the screen.}
\label{fig:abstraction}
\end{figure}

The conventional approach to QoE modeling is based on {\em human engineered features},
including quality metrics such as video rebuffering events, page load time, etc., and
other features such as screen size, genre of the video/website, etc..
Over the decades, a large body of research has expanded the set of features and
quality metrics, but recently there has been a dramatically acceleration
with not only more features but more finer-grained features handcrafted for each
application, each content type, and even each individual piece of content (\ie a
specific video or web page). Indeed, the march of QoE feature engineering is in full
swing, fueled by new applications (\eg interactive live video) on new
devices (\eg panoramic headsets) with new application behaviors (\eg new online
advertising methods). For example, recent work predicts video QoE using 
275 features~\cite{wassermann2019machine} and complex machine-learning
models~\cite{cardenas2019application}.
But the improvement in performance is not commensurate with this complexity.

The root of this complexity apocalypse, we argue, is that quality metrics and features act
as an ``information bottleneck'' that reduces QoE perception to a list of values, when in
fact the way users perceive QoE is more complex than what a (potentially long) list of
values can capture.
For instance, a user watching a video rarely perceives its quality by consciously counting
how long each stall lasts; instead, the video stalls (and other quality incidents)
influence the user's full viewing experience, including how much a
stall disturbs their engagement with the video content (\eg if it occurs during an
important like a goal in a live sports game).  By reducing QoE to a
handful of metrics, it would be difficult to characterize the impact of such quality incidents on the user's
full experience. In other words, {\em the problem with today's QoE models is not that they
need more features or complexity; it is that the feature-based abstraction of quality is a
mismatch for capturing user quality perception.}

In this paper, we propose a new abstraction for QoE modeling called {\em \video}---a video
stream that records all of the pixels displayed on the screen over time as seen by the
user, including the activities of all visible windows/frames. For example, it may
capture streaming video being played back by a viewer, or a sequence of web objects being rendered
by a browser.
Figure~\ref{fig:abstraction} contrasts the traditional QoE abstraction based on quality
metrics with a \video.
A \video is fundamentally distinct from the static content of an application (\eg the raw
video or web page content): it captures the rendering of the content on the screen after
compression, reordering, and any other effects of the application and network protocols
have taken effect.
The abstraction of \video enjoys two unique advantages over the traditional feature-based
abstraction:
\begin{packeditemize}
\item \Video captures the {\em full visual experience} of a user, which encompasses
the information captured by existing quality metrics, future ones we may discover, and
others way may never discover.

\item \Video applies to all Internet applications, because users experience
these applications by viewing pixels on a screen. Thus it is a {\em unifying} abstraction
that could potentially lead to unified QoE models across applications. 
\end{packeditemize}

Now, it may seem counter-intuitive that we address the high complexity of QoE modeling by
using a seemingly more complex abstraction. However, recent trends give us reasons to be
optimistic.
Computer vision has been revolutionized by the transition from traditional feature-based
models to far more accurate and general deep learning models, and the key enabling idea is
to learn useful representations directly from raw images, rather than handcrafted
features.
Inspired by this success, we believe a similar transformative approach can be applied to
QoE modeling, especially since QoE perception and computer vision share the visual
perception process. Although computer vision techniques and deep learning have been used
for QoE optimization, they have been used within the framework of feature-based QoE
modeling, \eg modeling the relationship between quality metrics and QoE
(\eg~\cite{yue2019deep,zhang2018deepqoe}) or deriving quality metrics from the static content
(\eg~\cite{liu2018end}).
We believe the time has come for a redefinition of the QoE abstraction, driven by
both application ``pulls'' (\eg user experience as the key driver) and technology
``pushes'' (\eg advances in computer vision).

\tightsection{Why QoE Modeling Matters}

We believe that accurate QoE modeling is the key to achieving higher QoE
in the face of limited network resources.

\tightsubsection{The QoE-resource trade-off}

Today's Internet users have much higher expectations for application quality than a few
years ago. As more applications move to mobile interfaces, users are becoming increasingly
impatient and sensitive to sub-second increases in page load time~\cite{Google-mobile}.
The surge of live videos (\eg~\cite{interactive-videos}) has shifted people's
perception of Internet videos from on-demand streaming to real-time interaction with a
massive, live audience. This growing demand for low delays is being met with a craving for
ultra high-quality content. With mainstream content providers and websites offering
more videos in 4K or higher resolutions~\cite{4k-akamai,4k-tv}, Internet video viewers
today demand much higher quality than ever before.

At the same, network resources are growing not as fast and not as evenly.
The disparity of broadband network access at home is a widespread phenomenon, even in the
US~\cite{broadband-disparity}.
The gap between limited network resources and the quest for higher application quality
underscore the need for new techniques that achieve {\em better QoE-resource tradeoffs}:
either achieving higher QoE with the same resources or reducing resource demands
without hurting QoE.



\tightsubsection{Accurate QoE modeling is the key}

Applications use a wide range of control algorithms to optimize the QoE of Internet video
(\eg~\cite{yan2020learning,akhtar2018oboe,yin2015control,mao2017neural}) and web services
(\eg~\cite{klotski,webgaze,wang2016speeding}) under dynamic availability of network
resources; accurate QoE modeling is key to the success of most of these techniques.
At a high level, a QoE-optimizing control algorithm can be framed as choosing the optimal
control action $a^*$ (\eg selecting the video bitrate, prioritizing web objects in a page,
etc.) from an action space that maximizes the expected QoE: $a^* = \textrm{argmax}_{a\in
A}(\hat{Q}(a,\hat{r}))$, where $\hat{r}$ estimates the available network resources
(bandwidth, latency, etc.) and $\hat{Q}(a,\hat{r})$ is the estimated QoE when taking
action $a$ under $\hat{r}$. Although this equation formulates a single-step optimization,
control algorithms typically optimize a longer-term QoE objective, which has important
implications for QoE modeling, but the idea is the same.

A considerable amount of research has focused on making accurate predictions of $\hat{r}$
to improve QoE under a given $\hat{Q}$.
We argue that accurate modeling of QoE ($\hat{Q}$) is at least as important as accurate
predictions of network resources. In particular, the QoE model fundamentally limits the
scope for improvement of all control algorithms for three reasons:

\begin{packeditemize}

\item {\em Balancing conflicting objectives:}
Applications are often faced with conflicting quality objectives.
For instance, video QoE can be improved by increasing average bitrate, avoiding bitrate
switches and rebuffering (stalls), and reducing start-up delay, but maximizing 
bitrate and minimizing join time are often in conflict, especially for short videos, and
minimizing bitrate switches often conflicts with maximizing bitrate and reducing
rebuffering~\cite{balachandran2013developing}, especially when bandwidth varies a lot.
In such settings, an accurate QoE model is crucial for adaptive bitrate (ABR) algorithms 
to strike a good balance among the objectives.

\item {\em Identifying which actions matter:}
Not all quality improvements lead to higher QoE, \eg because users have limited cognitive
capacity to perceive the change. 
For instance, when the page load time is below or above certain thresholds, it
may be too fast or too slow for users to experience a quality
difference~\cite{zhang2019e2e}.
Knowing exactly when quality improvements have a diminishing impact on QoE is critical to
achieving high QoE with minimum resources.

\item {\em Limiting action granularity:}
Finally, finer-grained QoE models allow finer-grained adaptation actions.
For instance, it has been shown that the same video bitrate leads to different
user-perceived quality depending on the video
content~\cite{nathan2019end,vmaf,guan2019pano}.
If a QoE model is agnostic to the perceptual quality of each bitrate on each video chunk,
then the ABR algorithm will not be able to raise/lower the bitrate of the chunks
where it matters more/less, thus missing opportunities to improve QoE or save bandwidth.

\end{packeditemize}



\begin{figure}[t]
\centering
\includegraphics[width=0.95\columnwidth]{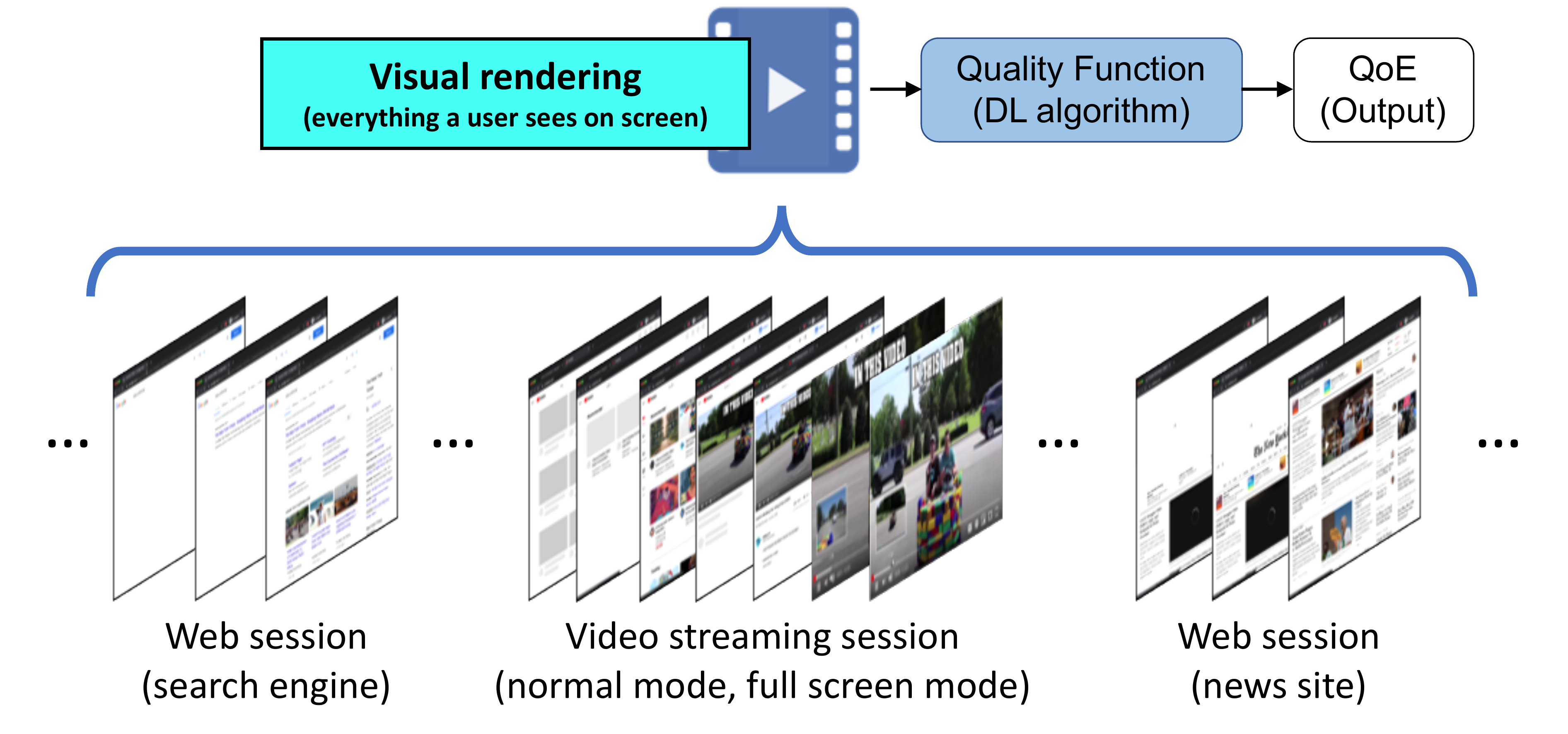}
\tightcaption{An illustration of \video and QoE modeling based on \videos.
}
\label{fig:rendering}
\end{figure}

\tightsection{The Case for Visual Rendering}

We introduce \video as a new abstraction for modeling QoE, outline its
benefits and potential over current approaches, and discuss similar notions in prior
work.

\tightsubsection{Today: Quality metrics as features}
\label{subsec:metrics}

Although a plethora of techniques exist for modeling the QoE of Internet applications,
they all share the same high-level {\em feature-based} approach.
They first extract handcrafted features or {\em quality metrics}, and
then build {\em quality functions} to model the relationship between these features and
user QoE. Each quality metric is crafted to capture some aspect of application
quality that might affect user QoE.

Quality metrics are widely used in industry as key performance indicators for optimizing
QoE, because they show stronger correlations with user ratings and engagement than
traditional packet/flow-level performance metrics.
For instance, video streaming QoE is modeled using metrics of visual quality of the
rendered frames (\eg SSIM~\cite{ssim}), quality stability (\eg number of bitrate
switches), and smoothness (\eg rebuffering events~\cite{dobrian2011understanding}). The
quality functions range from linear combinations of these metrics~\cite{yin2015control} to
deep learning models~\cite{cardenas2019application}.
Similarly, web QoE is modeled by variants of page load time (\eg time-to-first-byte)
to capture the impact of object loading progress on user
QoE~\cite{bocchi2016measuring,brutlag2011above,da2018narrowing}.



\tightsubsection{A new abstraction: Visual rendering}

We explore a new approach: instead of modeling QoE as a function of handcrafted features,
we instead model it directly from the pixels a user sees on the screen over time,
including the activities of all visible windows. In a video streaming application this
could be the frame-by-frame playback of a streamed video; in a web service this could be
the visual sequence of web objects loading in a web browser. We call this a {\em \video}.
Figure~\ref{fig:rendering} illustrates an example of a \video from a web browser, in which
it first loads a web page (search engine), then streams a video, and then loads another
web page (news).  Here, we assume the browser covers the full screen, but in general
a \video may include multiple windows.

\myparaq{What is captured (and not captured)}
Intuitively, a \video represents {\em all visual input to the visual perception process}.
It thus captures both the spatial experience (objects appearing on the screen at the same
time) as well as the temporal experience (dynamic loading of objects or playback of a
video). That said, \video does not include any non-visual factors, such as audio
information (which is rarely integrated by QoE models, except those used in the acoustics
literature), or contextual information like the user's device, browser settings, etc.. 
However, as we later discuss, contextual information influences the 
\video seen by a user and hence must be accounted for in our modeling.



\begin{figure}[t]
\centering
\includegraphics[width=.55\columnwidth]{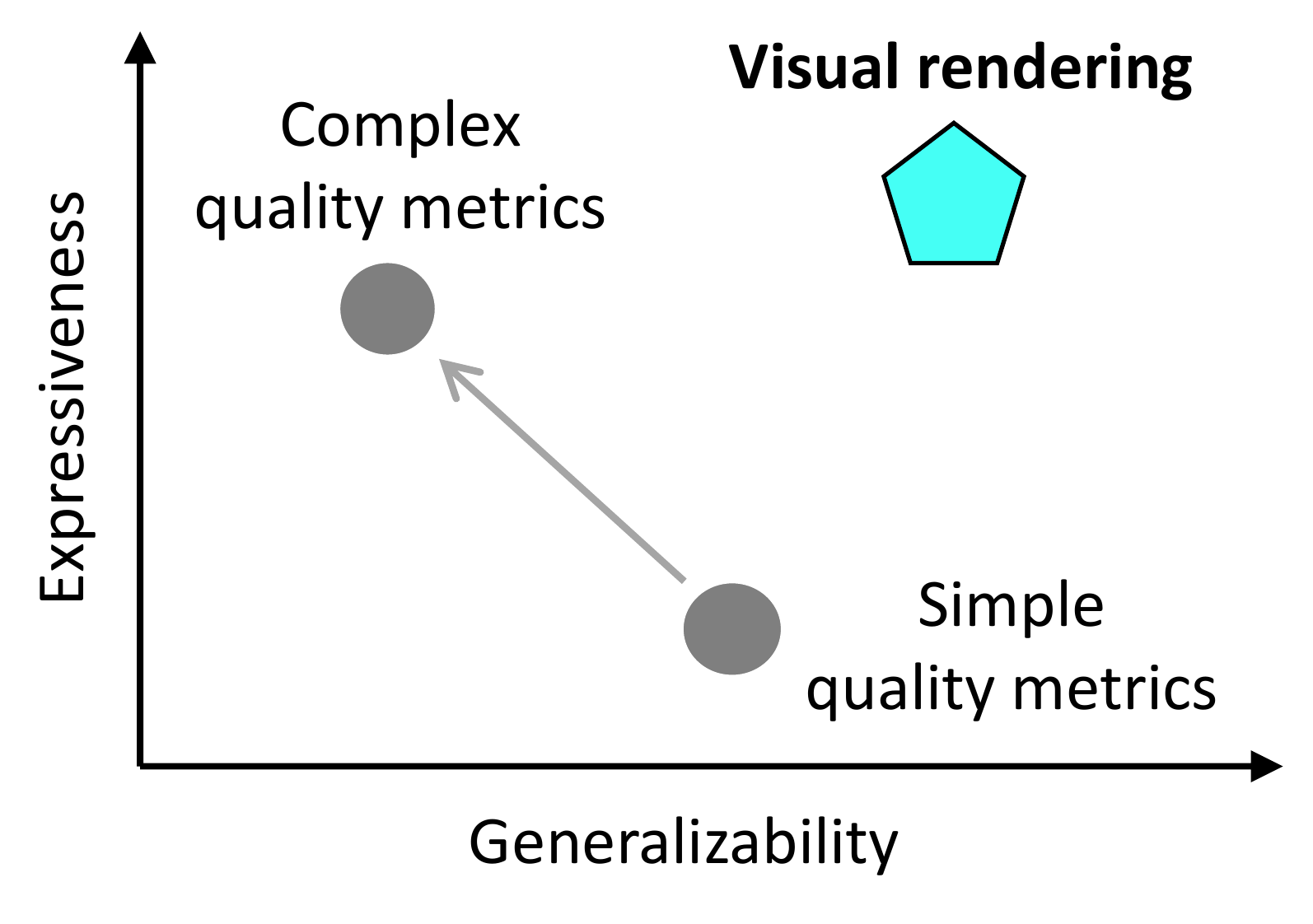}
\tightcaption{A qualitative comparison of traditional quality metrics and our
proposed abstraction of \video.}
\label{fig:tradeoffs}
\end{figure}


\tightsubsection{Benefits of \videos}

The key benefit of modeling QoE based on \video is that they are potentially
more {\em expressive} and more {\em generalizable} than feature-based QoE models (see
Figure~\ref{fig:tradeoffs}).

\subsubsection*{Expressiveness}

\begin{figure}[t]
\centering
\includegraphics[width=.95\columnwidth]{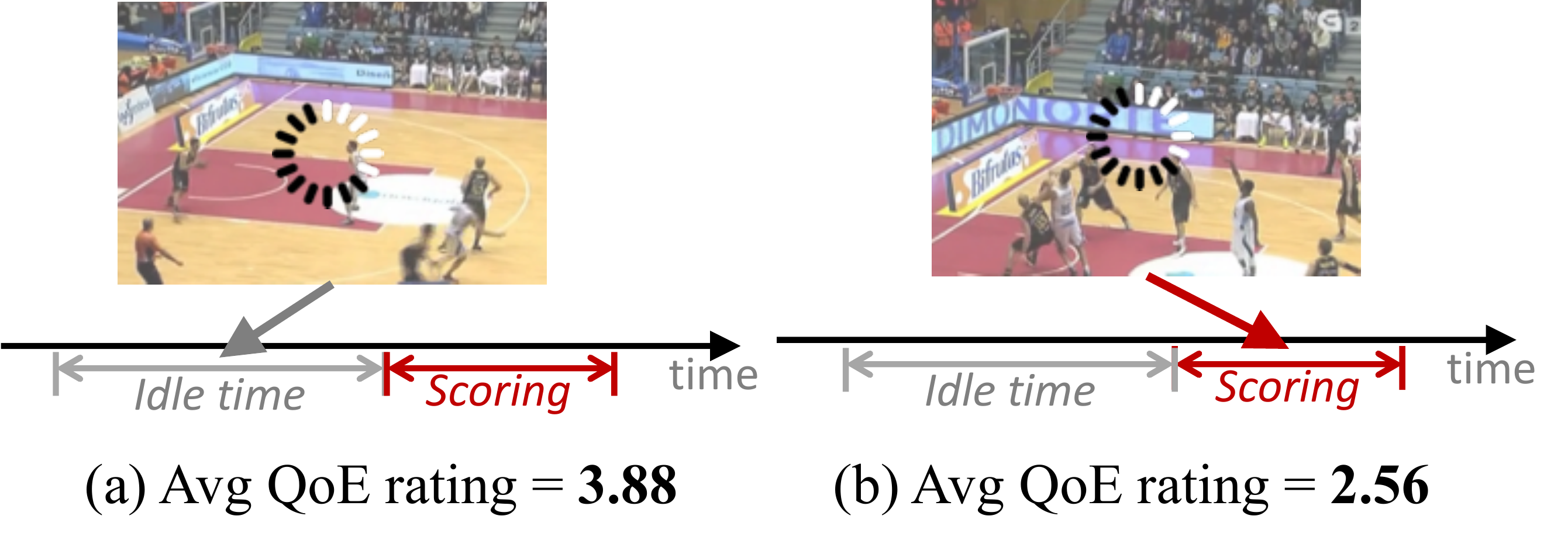}
\tightcaption{An MTurk study showing that videos with the same quality metrics can result
in significantly different user QoE depending on when a stall (indicated by the arrow) occurs.}
\label{fig:example}
\end{figure}

\begin{figure}[t]
      \begin{subfigure}[t]{0.325\linewidth}
         \includegraphics[width=1.0\linewidth]{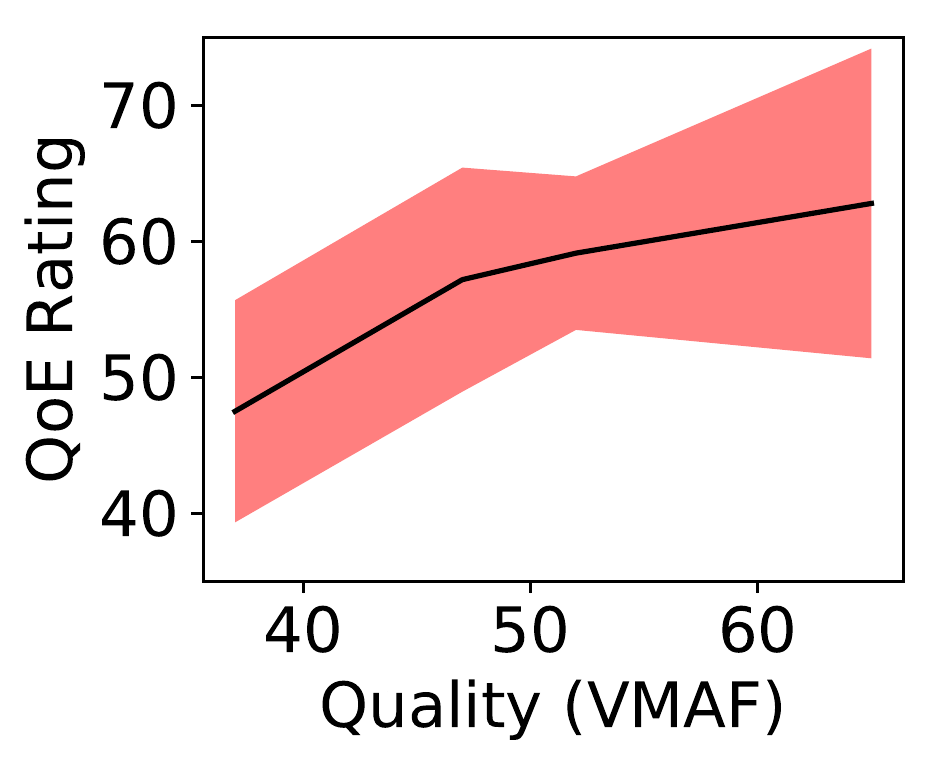}
        \caption{}
        \label{subfig:}
      \end{subfigure}
      \hfill
      \begin{subfigure}[t]{0.325\linewidth} 
         \includegraphics[width=1.0\linewidth]{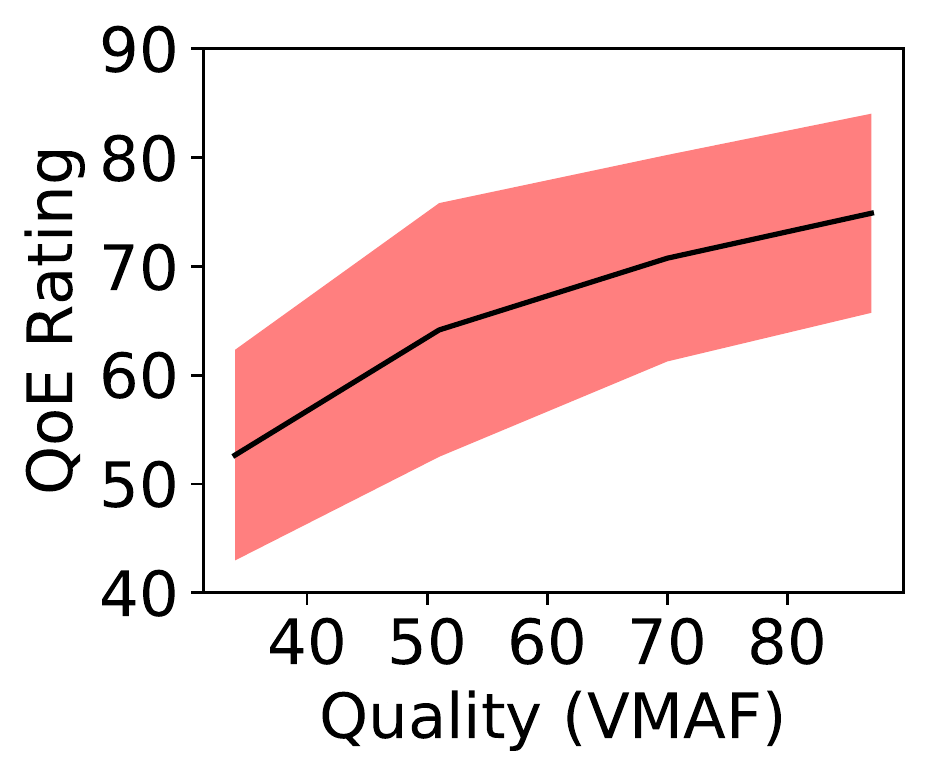}
         \caption{}
        \label{subfig:}
      \end{subfigure}
      \hfill
      \begin{subfigure}[t]{0.325\linewidth}
            \includegraphics[width=1.0\linewidth]{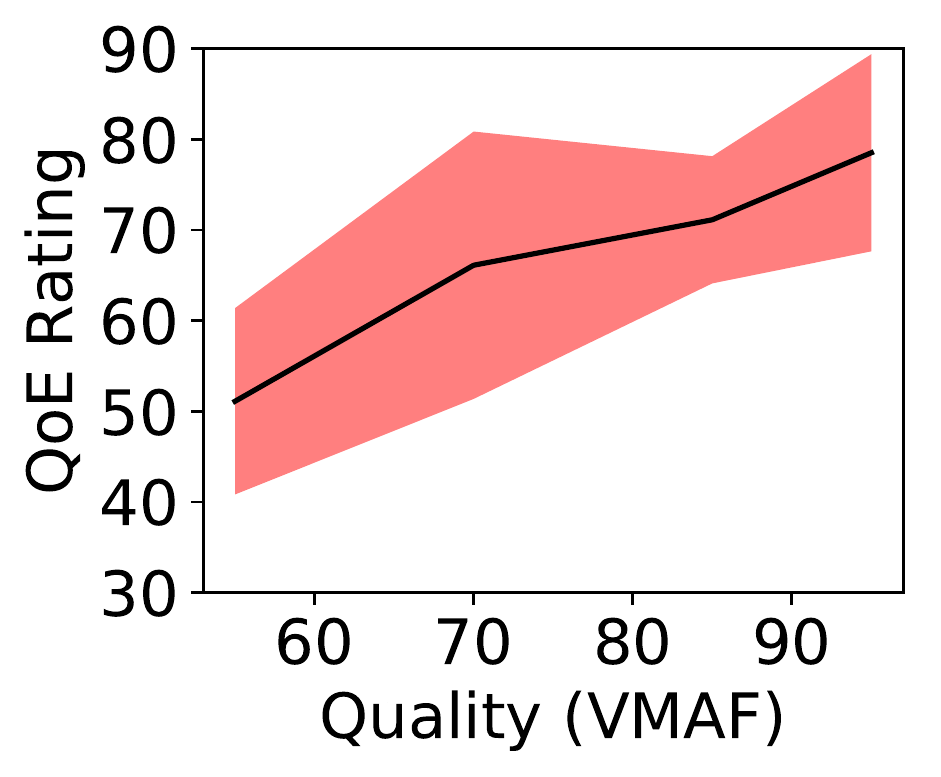}
            \caption{}
        \label{subfig:}
      \end{subfigure}
    \tightcaption{An MTurk study showing that visual quality metrics like VMAF cannot
    fully explain user QoE, as the high variance above shows (the lines show the mean and the belts show the stddev). Each mean and stddev
    is based on at least ten QoE ratings across the same set of 30 raters.}
    \label{fig:variance}
\end{figure}

Feature-based QoE models work well when the features (quality metrics) capture user QoE,
but QoE usually varies substantially even when we limit the values of key quality metrics
to a small range.
To show this, we create multiple videos and ask Amazon MTurkers to rate their QoE on a
scale of 1-5.
All videos show the {\em same} content at the {\em same} bitrate and include the {\em
same} half-second rebuffering stall, and are rated by the {\em same} MTurkers; the only
difference is when the stall occurs.
Most QoE models would predict the same QoE for all videos, but we observe systematic
differences in the mean rating of each video, as shown in Figure~\ref{fig:example}. This
is because QoE ratings drop sharply when the stall occurs at a critical moment in the
video.
A similar effect occurs even for quality metrics that are content aware.
For example, VMAF~\cite{vmaf} is a visual quality metric that gives lower QoE estimates if
a bitrate drop occurs when frame pixels are more ``complex''.
We run a large-scale analysis on a public video QoE dataset~\cite{duanmu2018sqoe} and
ask MTurkers to rate videos with similar (to within 5\%) rebuffering time,
number of bitrate switches, and VMAF scores. Figure~\ref{fig:variance} shows the means and
variances of QoE ratings (on a scale of 0-100) against VMAF score. We see that the
variances are consistently more significant than the differences in mean QoE rating due to
higher VMAF scores, which means that VMAF cannot adequately predict user QoE.
Although the variances might be explained by finer-grained quality metrics, finding all
such metrics is infeasible, and prediction errors like this are common in even
state-of-the-art video/web QoE models.

\ignore{
\jc{maybe pick on of the two examples if short on space}
}

\mypara{Why \videos might be expressive}
In contrast, \videos by definition preserve all visual information that affects a user's
experience, in both video and web-based applications.
This not only includes all the information needed to derive existing quality metrics (\eg
rebuffering time and page loading delay), but also preserves other information that might
affect QoE, including factors that we have yet to discover.
These factors include how content and application quality affect a user's perceived QoE.
For instance, the QoE variance in Figure~\ref{fig:example} and~\ref{fig:variance} may
be caused by how visual content affects the relationship between application
quality and QoE, which is captured by a \video.

\ignore{
\footnote{We
recognize that a \video alone cannot capture certain visual quality metrics that compares
original content with encoded one, but there are also no-reference video quality
metric~\cite{gu2016no}, and it is also possible to augment \video with the original
content, just as prior QoE models.}}


\Videos also include information that can potentiall help model how new adaptation actions
affect QoE.  For example, consider a new video adaptation action where the player
(slightly) slows down the playback of a video while replenishing the buffer, in order to
avoid more abrupt stalls. Traditional feature-based QoE models cannot
capture this effect because no feature or quality metric is designed for this action.
In contrast, \videos naturally include all temporal information, including the slowdown of
the video.

\subsubsection*{Generalizability}

Traditional feature-based QoE models trade complexity for higher accuracy.
A case in point are the quality metrics used for rebuffering: initially, its impact on
video QoE was measured using the rebuffering ``ratio'' (the fraction of time spent in
stalls during a video session)~\cite{dobrian2011understanding}, but more complex metrics
emerged over time, capturing factors such as the relationships between rebuffering stalls
(\eg length distribution and memory effect~\cite{eswara2019streaming,duanmu2018quality})
and the differences in its effect on the QoE of live videos versus on-demand
videos~\cite{balachandran2013developing}.
As QoE models become more complex and fine-grained, however, they also become harder to
generalize.
For instance, traditional QoE models might be able to explain the QoE variance in
Figure~\ref{fig:example} if they are customized to each video; but such per-video QoE
models do not generalize and are prohibitively costly to create
(\S\ref{subsec:training-data}).
Similarly, web QoE models that are specialized to a web page can predict QoE much more
accurately than a one-size-fits-all QoE model~\cite{da2018narrowing}, but creating such
per-page QoE models also faces a scalability problem, especially since content is
continuously changing.

\begin{figure}[t]
\centering
\includegraphics[width=.95\columnwidth]{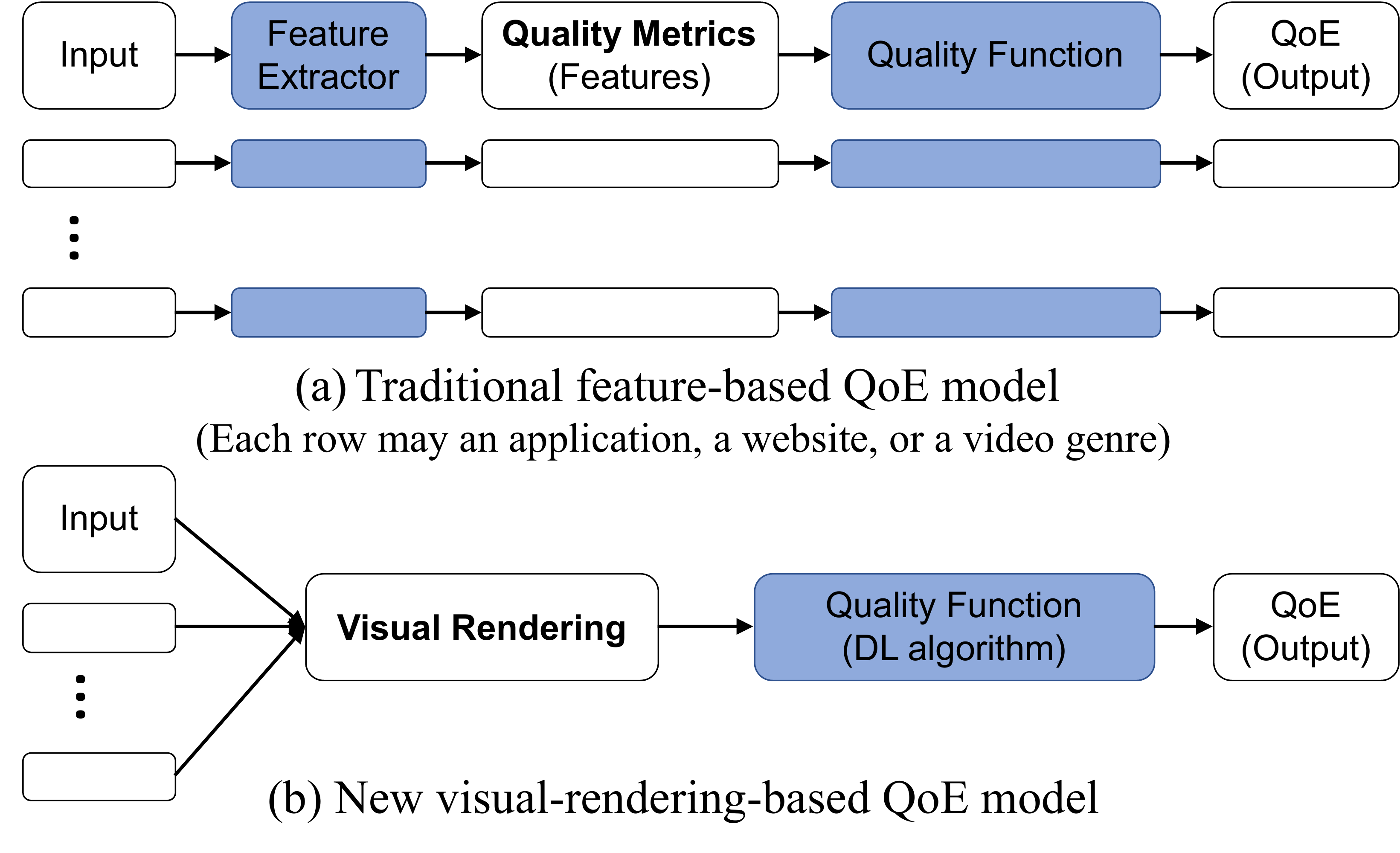}
\tightcaption{\Videos may offer a more generalizable approach to QoE modeling, by
providing a unifying input to deep learning models that directly predict user QoE.
}
\label{fig:contrast}
\end{figure}

\mypara{Why \videos might be generalizable}
It is difficult to say upfront, but we have reasons to believe that a QoE model based on
\video will generalize. There is a striking analogy between Internet QoE research and
computer vision research in the pre-deep-learning era. Back then, each computer vision
task (\eg object detection, gesture detection, segmentation) had a separate literature
that developed handcrafted features customized for the task.
The success of deep learning in computer vision is not only that it provides more
accurate models, but also that it provides a generalizable approach.
Deep learning models take the raw pixels as input and are trained ``end-to-end''
with minimal domain-expert intervention.
Moreover, deep learning models for different computer vision tasks often share the
same convolutional layers (\eg ResNet) as common feature extractors, which are more
expressive than the best handcrafted features.
We speculate that building end-to-end deep learning models directly from \videos might
lead to a more generalizable approach to QoE modeling than relying on handcrafted
features/models, as Figure~\ref{fig:contrast} illustrates.

%
%
%
%

\tightsubsection{Similar concepts in prior work}

Several concepts from prior work are closely related to \videos, but do not take them to their logical extreme. 

\mypara{Gaze tracking/prediction}
WebGaze~\cite{webgaze} shares with us the insight that user gaze varies with the dynamic web loading process, implying that web QoE is influenced by the gaze trajectory in addition to traditional page load time metrics.
In particular, WebGaze tracks gaze while a web loading process is replayed, which is similar to a \video.
Some follow-up work automatically derives user gaze from web content (\eg~\cite{kelton2019reading,vidyapu2019quantitative}), and similar techniques are also used to track user saliency in panoramic videos (\eg~\cite{nguyen2018your,zhang2018saliency}).
However, these efforts use gaze or saliency as another feature in traditional QoE models (\eg to reweight web objects or video pixels/chunks).

\mypara{Eliciting QoE feedback} 
EYEORG~\cite{eyeorg} uses recorded videos to elicit user ratings (QoE) for video streaming and web services. 
They do this because users may have different network connectivities, so rather than letting them stream the videos or load the web pages, they show users pre-recorded videos of a video session or web page loading process.
Though the idea of showing recorded videos resembles the concept of \videos, EYEORG and others~\cite{wu2013crowdsourcing} still model QoE as a function of pre-determined quality metrics.

\tightsection{Architecting for Visual Rendering}

So far we have seen that the abstraction of \videos~{\em could} lead to more expressive and generalizable QoE models.
In this section, we discuss the technical challenges to realizing this ideal.
At this stage of our research, we do not yet know if the advantages of \videos will outweigh the challenges.
We recognize that our vision for re-architecting QoE frameworks is broader than what we can accomplish alone. 
By outlining a specific research agenda, we hope to spark discussions and efforts from the networking, multimedia and computer vision communities.

\begin{figure}[t]
\centering
\includegraphics[width=.99\columnwidth]{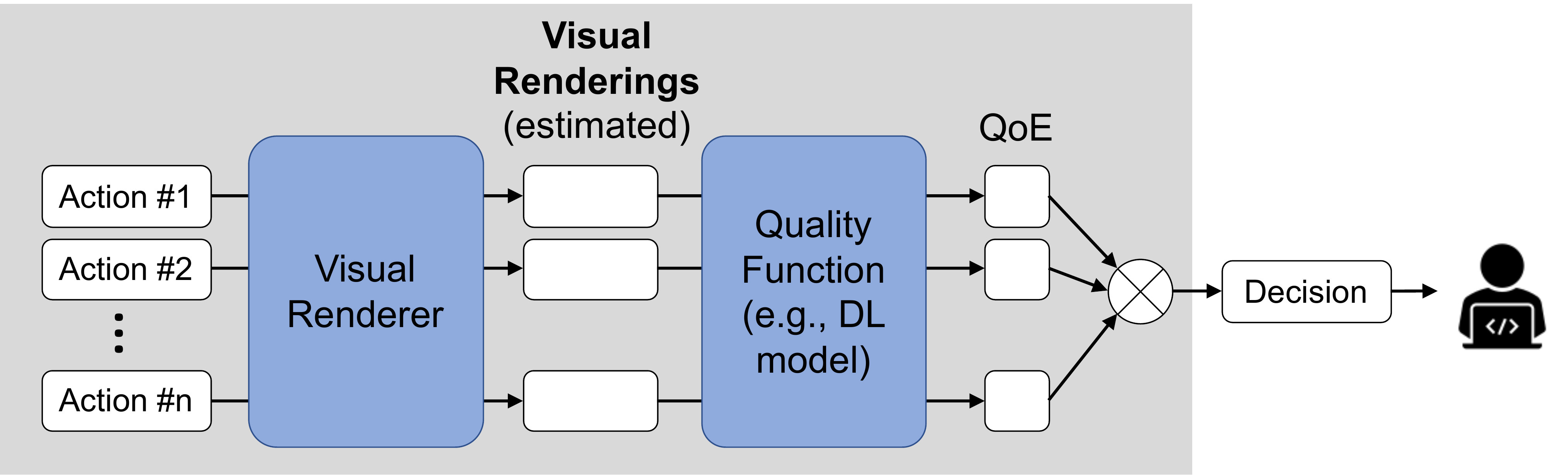}
\vspace{0.1cm}
\tightcaption{A framework for QoE optimization based on \videos and deep learning models.
}
\label{fig:workflow}
\end{figure}

\mypara{Optimization architecture}
Figure~\ref{fig:workflow} depicts a logical view of a QoE optimization framework based on \video. 
It applies two components to each adaptation action: 
\begin{packeditemize}
\item A {\em \videor (4.1)} first infers the \video of the action, and 
\item A \video-based {\em QoE model (4.2)} then predicts the QoE of a given \video.
\end{packeditemize}
Finally, we pick the action that achieves the best QoE.
The \video of an action may also include the recent \videos up to this point, since QoE is often dependent on the content and the user's QoE in recent history. 
This can be done implicitly by the emulator or with help of the client-side browser.

\tightsubsection{\Video emulation}

Inferring a \video from an action in real-time is a formidable task, because the \video may depend on the specific content being shown as well as the context of the user's video or web session. In an ideal world, we would freeze time, take the action in a parallel world for the same user, capture the \video it results in, and feed that to our QoE model. Since this is not possible, we must find an alternative approach.

\mypara{Leveraging existing testing infrastructure}
Web content providers rely on extensive testing infrastructure to evaluate their application protocols and control algorithms, including automated unit tests, A/B testing frameworks, human testers, and others. Some of these testing environments emulate the experience of streaming a video or loading a web page, providing an ideal opportunity to capture a \video. However, even if we are able to tap into this infrastructure to enumerate all possible \videos that result from  the adaptation actions of an application, we still face two serious challenges to making this viable:
\begin{packeditemize}
\item {\em Diverse clients:} The \video experienced by a user is influenced by several contextual factors such as the user's device, available bandwidth, browser settings, etc..

\item {\em Real-time decisions:} There is very little time between when a user request arrives and when an adaptation action must be taken to deliver content to the user.
\end{packeditemize}

These challenges imply that a \video must be contextualized to the user in real-time. Since creating a \video from scratch is not feasible in real-time, and since offline-enumerated \videos (such as the ones mentioned above) are not contextualized to the user, we propose a compromise: {\em parameterized \videos}. That is, we enumerate parameterized \videos offline that can take contextual factors as input online and quickly specialize the \video to those factors. Although this is still a difficult task, consider the following examples. If we record a \video assuming a particular network bandwidth, we can emulate other network bandwidths by simply speeding up/slowing down the \video. Similarly, if we record timings in the \video of when distinct web objects are loaded, we might be able to speed up/slow down specific object loading events, or even rearrange the load order (with additional video editing effort).



\tightsubsection{\Video-based QoE modeling}
\label{subsec:modeling}


\subsubsection*{Designing a \video-based QoE function}
We have two intuitive reasons to posit that a general \video-based QoE model is plausible.
First, from a cognitive perspective, the perception of streaming video and web browsing involve the same psychophysical process.
Second, \videos enable us to harness the power of deep-learning-based computer vision, which also models human perception.
We elaborate on both aspects below.

\mypara{Drawing ideas from cognitive visual perception}
Visual perception is a primary focus of cognitive research.
It aims to reveal the general psychophysical process behind {\em all} visual perception activities, which include web browsing and watching videos.
There are two key concepts: {\em expectation}, which describes how prior experience affects the perception of visual stimuli, and {\em attention}, which influences the neuronal representation of current visual stimuli~\cite{gordon2019expectation}.

There is a striking parallel between these two concepts and how application quality affects QoE in networking research.
For instance, a video rebuffering event (stall) is a violation of the expectation since the user expects the video to continue playing.
Similarly, fast loading of a web page means higher QoE, because it meets the expectation of a user when a link is clicked.
A user's expectation of application quality is also shaped by the quality of recent web/video sessions~\cite{hossfeld2011memory,duanmu2018quality}, which has been studied under the framework of cognitive biases.
Similarly, models of human visual attention are increasingly used in \vr videos~\cite{ozcinar2019visual,nguyen2018your} and 
recently in web optimization~\cite{kelton2019reading,vidyapu2019quantitative,percival}.
In short, we posit that {\em using the concepts of expectation and attention, high QoE can be interpreted as having less violation of expectation within the region of attention.}



\mypara{Drawing ideas from computer vision}
While the visual perception literature provides a useful framework for understanding QoE, we still need to automatically infer attention and expectation.
This is where computer vision might provide useful building blocks.
In the interest of space, we only highlight the three most relevant topics.
(1) {\em Visual attention (saliency) detection}~\cite{wang2019salient,wang2017deep} uses convolutional models to reason about the spatial structures that influence the distribution of human visual attention. 
(2) {\em Video summarization} (and {\em highlight detection})~\cite{zhang2016video} uses recurrent models to learn the temporal patterns in a video and when users will pay more attention to high-level incidents.
(3) {\em Video prediction} predicts future video frames based on the previous ones, which helps to model user expectation of the content.

\mypara{Open questions}
Despite the apparent congruity between QoE and computer vision, their mismatch is also evident. 
\begin{packeditemize}
\item {\em What should the QoE model look like?} 
We can use mature techniques such as the attention mechanism to model attention and recurrent models to learn temporal patterns in a \video, but combining them is challenging.
One idea is to merge them similar to how computer vision models and natural language models are combined to perform high-level tasks such as visual question-answering.
We also speculate that the QoE model of one application could be fine-tuned to serve other applications by transfer learning, via  requires less training data.


\item {\em \Videos are not ``natural'' videos:} 
Computer vision works well with natural images/videos that do not have artificial glitches (\eg video rebuffering or bitrate switches)
that influence QoE.
For instance, quality incidents such as a video stall or bitrate switch can affect user attention/expectation (as observed in~\cite{webgaze}) but they are rarely modeled in computer vision.
\end{packeditemize}


%
%

\subsubsection*{Creating new QoE datasets for training the model}
\label{subsec:training-data}

\mypara{Existing datasets are inadequate}
Training a QoE model requires an annotated visual rendering dataset that covers many combinations of content and quality incidents. 
Unfortunately, existing QoE datasets have limited variability of the video/web content.
For instance, popular video QoE datasets include only a handful of videos (20 or less~\cite{duanmu2018quality,netflix-dataset,eyeorg}), in part because QoE data collection can be frustratingly slow and expensive---to test one video content, researchers need to recruit tens of participants and let each of them watch the same video rendered with different quality incidents.


\mypara{The wisdom of crowd}
A potential solution is to leverage commercial crowdsourcing platforms such as Amazon Mechanical Turk~\cite{mturk}. 
For its short response times, auto scaling, and reasonable pricing, crowdsourcing is a promising alternative to lab studies for QoE annotation~\cite{zhang2019e2e,wu2013crowdsourcing}.
That said, existing use of crowdsourcing platforms only models specific relationships between quality metrics/features and QoE. 


\mypara{Open questions}
There are two key questions:

\begin{packeditemize}
\item {\em How to create a \video-based QoE dataset?}
One idea is to draw from popular content (\eg the Alexa top-1000 web sites), but popularity does not necessarily mean adequate diversity.
Alternatively, one can sample across many content genres similar to how ImageNet compiles images of different objects from each class.

\item {\em Other sources of data?}
We recognize that a scaled-down version of the envisioned dataset can be built by a content provider (\eg Netflix or Google).   
A content provider can passively monitor \videos seen by its users and label each \video with the user engagement (how long a user watches a video or stays on the web site) as the QoE.
This process can easily generate a large amount of annotated data, but the content could be biased.
\end{packeditemize}

We do not claim the ideas outlined here are the only (or optimal) way of building the envisioned QoE model. Instead, we hope they inspire more ideas and research.

\tightsection{A New Frontier for ML in Networking}
\label{sec:conclusion}

Machine learning is increasingly used in networking, but so far it has largely been a  ``solver'' of complex control problems such as scheduling, bitrate adaptation, and resource selection.
The abstraction of \video creates a new frontier for harnessing the power of deep learning, which  revolutionized computer vision and may similarly transform user-facing applications and Internet QoE.
We believe the confluence of trends---user QoE as the key driver and recent advances in computer vision---make now the right time to explore this frontier.






\bibliographystyle{abbrv} 
\begin{small}
\bibliography{reference}
\end{small}












\end{document}